\documentclass[
  journal=pasa,
  manuscript=research-paper, 
  year=2020,
  volume=37,
]{cup-journal}

\usepackage{microtype,siunitx,booktabs}

\usepackage{graphicx}
\usepackage{amsmath}	
\usepackage{amssymb}	
\usepackage{multicol}       
\usepackage{bm}		
\usepackage{pdflscape}	
\usepackage{hyperref} 
\usepackage[T1]{fontenc}
\usepackage{ae,aecompl}
\usepackage{xcolor} 
\usepackage{booktabs}
\usepackage{nicematrix}
\title[Morphological Identification using Weak Labels]{Deep Learning for Morphological Identification of Extended Radio Galaxies using Weak Labels}

\author{Nikhel Gupta$^{1}$}
\author{Zeeshan Hayder$^{2}$} 
\author{Ray P. Norris$^{3,4}$} 
\author{Minh Huynh$^{1,5}$}
\author{Lars Petersson$^{2}$}
\author{X. Rosalind Wang$^{3}$}
\author{Heinz Andernach$^{6}$}
\author{B\"arbel S. Koribalski$^{4,3}$}
\author{Miranda Yew$^{3}$}
\author{Evan J. Crawford$^{3}$}
\email[Nikhel Gupta]{Nikhel.Gupta@csiro.au}
\affiliation{
$^1$ CSIRO Space \& Astronomy, PO Box 1130, Bentley WA 6102, Australia \\
$^2$ CSIRO Data61, Black Mountain ACT 2601, Australia \\
$^3$ Western Sydney University, Locked Bag 1797, Penrith, NSW 2751, Australia \\
$^4$ CSIRO Space \& Astronomy, P.O. Box 76, Epping, NSW 1710, Australia \\
$^5$ International Centre for Radio Astronomy Research (ICRAR), M468, The University of Western Australia, 35 Stirling Highway, Crawley, WA 6009, Australia \\
$^6$ Th\"uringer Landessternwarte, Sternwarte 5, D-07778 Tautenburg, Germany.
Permanent address: Depto.\ de Astronom\'{i}a, DCNE, Universidad de Guanajuato, Cj\'on.\ de Jalisco s/n, Guanajuato, CP 36023, Mexico 
}

\received {dd Mmm YYYY}
\revised  {dd Mmm YYYY}
\accepted {dd Mmm YYYY}
\published{dd Mmm YYYY}


\usepackage{amsmath,amsfonts,amssymb}
\usepackage{color}
\usepackage{mathrsfs}
\usepackage{tabularx}
\usepackage{floatrow}
\usepackage{float}

\definecolor{ored}{rgb}{1.00,0.27,0.00}
\definecolor{mygreen}{rgb}{0.2,0.7,0.2}

\definecolor{Gray}{gray}{0.5}
\definecolor{LightCyan}{rgb}{0.88,1,1}

\def \BE{\begin{equation}}
\def \EE{\end{equation}}	
\def \BC{\begin{center}}
\def \EC{\end{center}}
\def \BEA{\begin{eqnarray}}
\def \EEA{\end{eqnarray}}

\def \SIGMA8{\sigma_{8}}

\usepackage{cellspace}
\begin{document}\sloppy\sloppypar\raggedbottom\frenchspacing

\begin{abstract}
The present work discusses the use of a weakly-supervised deep learning algorithm that reduces the cost of labelling pixel-level masks for complex radio galaxies with multiple components. 
The algorithm is trained on weak class-level labels of radio galaxies to get class activation maps (CAMs).
The CAMs are further refined using an inter-pixel relations network (IRNet) to get instance segmentation masks over radio galaxies and the positions of their infrared hosts.
We use data from the Australian Square Kilometre Array Pathfinder (ASKAP) telescope, specifically the Evolutionary Map of the Universe (EMU) Pilot Survey, which covered a sky area of 270 square degrees with an RMS sensitivity of 25-35 $\mu$Jy/beam. 
We demonstrate that weakly-supervised deep learning algorithms can achieve high accuracy in predicting pixel-level information, including masks for the extended radio emission encapsulating all galaxy components and the positions of the infrared host galaxies. 
We evaluate the performance of our method using mean Average Precision (mAP) across multiple classes at a standard intersection over union (IoU) threshold of 0.5.
We show that the model achieves a mAP$_{50}$ of 67.5\% and 76.8\% for radio masks and infrared host positions, respectively.
The network architecture can be found at the following link: \url{https://github.com/Nikhel1/Gal-CAM}
\end{abstract}


\section{Introduction}
\label{SEC:Intro}
Recent technological advances in radio astronomy have enabled us to significantly reduce the integration time to survey the sky.
This has led to a new era in radio surveys, producing deep continuum images of hundreds of square degrees of radio sky at unprecedented depths.
Advanced radio interferometers like the Australian Square Kilometre Array Pathfinder \citep[ASKAP:][]{johnston07ASKAP,DeBoer09,hotan21}, 
the Murchison Widefield Array \citep[MWA:][]{tingay13,wayth18}, MeerKAT \citep{jonas16}, the Low Frequency
Array \citep[LOFAR:][]{vanharleem13} and the Karl G. Jansky Very Large Array \citep[JVLA:][]{perley11} are among the instruments being used for these surveys.
The upcoming multi-million catalogues of radio sources from these surveys will greatly enhance our knowledge of the Universe. 
To fully harness the potential of these surveys, there is a need to increase the efficiency of the data analysis techniques.

The improved resolution of wide-field radio surveys has led to the detection of increasingly more extended radio galaxies.
These galaxies may have multiple components with separate peak emissions.
While the related components of a galaxy may be connected to each other, for many galaxies these components may be separated by some distance.
A non-trivial task here is the process of morphological identification of all related components of a radio galaxy.
Currently, existing methods for component identification are primarily based on visual inspection which does not scale up to the massive data volumes of this decade.
Without redesigning these component identification methods, the radio source catalogues from the next-generation surveys may not be ideal for further scientific analysis.

Machine learning (ML) based on computer vision has become a powerful tool for extracting and modelling high-dimensional information from images in recent years. 
The available data for learning usually determines computer vision tasks, which are categorized into four types: supervised, self-supervised, semi-supervised, and weakly-supervised methods. 
In supervised learning, the model is trained on image-label pairs, where the labels correspond to the true and exact information necessary to train the model for a specific prediction. 
Recently, these ML methods have been used for morphological classifications of radio sources \citep[e.g.][]{lukic18, alger18,wu19,bowles20,viera21,becker21,brand23}, but these models cannot be used for learning without precise training labels. 
Labelling a large training dataset is expensive, and with multi-million radio detections in future surveys, obtaining true and exact labels for a reasonably large fraction is not feasible.

Self-supervised learning refers to the use of unsupervised methods to train models on the structure of the data without the need for explicit labels.
This approach has been used to identify new types of galaxies in radio surveys \citep[e.g.][]{galvin20, mostert21, gupta22}.
Since it does not rely on truth labels for training, self-supervised learning produces less reliable results compared to supervised learning methods that do rely on labelled data.
Semi-supervised learning can be used to address this issue by combining labelled and unlabelled data during training, as has been done for the classification of radio galaxies \citep{slijepcevic22}.
This approach may also reduce the amount of labelled data needed for training compared to supervised learning methods.
Alternatively, weakly supervised learning can use noisy, restricted, or indirect labels for the entire training data, which can reduce the labelling effort required for large datasets by using indirect information for each galaxy as a supervisory signal.

Various methods have been developed to classify radio galaxies and identify their associated components. 
Among the labelling efforts, image ``tags" or ``classes" are considered to be cost-effective, informative, and helpful in investigating and interpreting the underlying physics \citep[e.g.][]{rudnick21}. 
The Galaxy Zoo portal is one of the most effective efforts to tag images, which is publicly accessible for citizen scientists to participate in solving astronomical problems. 
Citizen scientists are typically individuals from society who have not received formal training in the scientific field.
An example relevant to the present work is the Radio Galaxy Zoo \citep[RGZ;][]{banfield16}.
In the case of RGZ, citizen scientists were tasked with providing tags to radio sources to determine the number of associated peaks and components for each galaxy. 
While these tags are helpful for supervised classification tasks \citep[e.g.][]{wu19}, they cannot be directly used to generate pixel-level information, such as segmentation masks for galaxies in radio images. 
It should be noted that such segmentation masks are crucial for identifying related radio components in the image.

This work presents a weakly-supervised instance segmentation method. 
In addition to the radio galaxy segmentation masks, we also detect positions of the infrared host galaxies using keypoint estimation.
Our weakly-supervised deep learning model achieves high accuracy in predicting pixel-level information, including masks for the extended radio emission encapsulating all galaxy components and the positions of the infrared host galaxies. 
We evaluate the performance of our deep learning algorithm to produce instance segmentation masks from weak labels. 
The positional accuracy of infrared hosts is also evaluated.

The paper is structured as follows.
In Section~\ref{SEC:Observations}, we describe the radio and infrared data and labels used for training and inference.
Section~\ref{SEC:method} is dedicated to the methods that include image pre-processing, a description of the deep learning pipeline, and details about the network training.
In Section~\ref{SEC:Results}, we present our findings from the trained network.
We summarise our findings in Section~\ref{SEC:conclusions} and provide directions for future work.

\section{Data}
\label{SEC:Observations}
This section presents the radio and infrared observations and the labels utilized for training and performing inference with a weakly-supervised ML network.
\subsection{Radio Observations with ASKAP}
\label{SEC:ASKAP}
Located at Inyarrimanha Ilgari Bundara, the Murchison Radio-astronomy Observatory (MRO), ASKAP is a radio telescope equipped with phased array feed \citep[PAF:][]{hay06} technology, which allows for a wide instantaneous field of view, resulting in high survey speed. 
ASKAP has 36 antennas with varying baselines, most located within a 2.3 km diameter and the outer six extending the baselines up to 6.4 km \citep{hotan21}. 
Recently, ASKAP completed the first all-sky Rapid ASKAP Continuum Survey \citep[RACS:][]{McConnell20} covering the entire sky south of Declination $+41^{\circ}$, with a median RMS of approximately 250 $\mu$Jy/beam. This survey has opened up the possibility of conducting subsequent deeper surveys using ASKAP.

A planned survey known as the Evolutionary Map of the Universe \citep[EMU;][]{norris11} aims to observe the entire Southern Sky, with an expected catalogue of up to 40 million sources \footnote{Forecast based on the allocated time for the EMU 5-year survey program (see https://www.atnf.csiro.au/projects/askap/commissioning\_update.html).}.
Taking steps in this direction, the EMU Pilot Survey \citep[EMU-PS:][]{norris21} was conducted towards the end of 2019. 
Covering a sky area of 270 deg$^2$ with $301^{\circ}< {\rm RA} < 336^{\circ}$ and $-63^{\circ}< {\rm Dec} < -48^{\circ}$, EMU-PS is composed of 10 tiles, with a total integration time of about 10 hours for each tile. 
It achieved an RMS sensitivity of $25-35~\mu$Jy/beam and a beamwidth of $13^{\prime\prime} \times 11^{\prime\prime}$ FWHM. 
The operating frequency range of EMU-PS is from 800 to 1088 MHz, centred at 944 MHz.
The ASKAPsoft pipeline \citep[][]{whiting17} was used to process the raw data. 
To avoid any potential systematic errors associated with automated source detectors, we rely for this on a catalogue of Double Radio sources Associated with Active Galactic Nuclei \citep[DRAGN; ][]{leahy93} wherein source identification is conducted solely through visual inspections.
The detailed process of identifying sources for the DRAGN catalogue will be outlined in \citep{yew22prep}. 
Here, we provide a summary of the procedure.

In the initial step, the EMU-PS radio image was quickly scanned for any obvious DRAGNs, and a circular region was defined around each of them. 
In the subsequent step, a thorough visual scan of the complete radio image was carried out by systematically examining each part, identifying and categorizing sources, and setting a circular region encompassing the diffuse radio emission for each. 
A third and final step involved an extensive scan to identify any sources that might have been missed and were not associated with a circular region.
After the identification process, each source's centroid position and approximate size were recorded in the catalogue.
Using the centroid position, we created $7^{\prime} \times 7^{\prime}$ cutouts for each DRAGN source. 
The resulting image is $210\times 210$ pixels in size, with a pixel size of $2^{\prime \prime}$.
The selected cutout size is appropriate as the largest source measures $6.5^{\prime}$.

\subsection{Infrared Observations}
\label{SEC:DES_SDSS}
Infrared images were obtained at the exact locations of the radio source cutouts.
We use the Wide-field Infrared Survey Explorer \citep[WISE;][]{wright10}, which is an infrared survey of the entire sky that detects radiation in the W1, W2, W3, and W4 bands corresponding to 3.4, 4.6, 12, and 22 $\mu$m wavelengths. 
In the present study, we focused only on the W1 band from the AllWISE \citep[][]{cutri13} dataset, which has a 5$\sigma$ point source sensitivity of 28 $\mu$Jy and angular resolution (FWHM of the major axes of the PSF) of $6.1^{\prime\prime}$. 

\subsection{Data Labelling}
\label{SEC:Labels}
The radio and infrared image cutouts contain four types of labels, which are class-level labels indicating the types of radio sources, and pixel-level labels that include bounding boxes placed at the locations of radio sources covering their connected or separate components, segmentation masks for radio sources, and positions of potential host galaxies in the infrared.
Our objective in this study is to evaluate the efficacy of a weakly-supervised deep learning algorithm trained with a restricted amount of labelled data. Therefore, we solely employ class-level labels during training, and pixel-level labels are exclusively utilized for model inference.
We will briefly overview the labelling procedure here, and a detailed discussion will be presented in \cite{yew22prep}.

\subsubsection{Infrared Hosts}
\label{SEC:IRhostLabels}
To determine the infrared host galaxy associated with each radio source, we performed a manual identification process by superimposing radio and infrared images.
The likely infrared host is typically near the geometric centre of radio sources. 
In the case of asymmetric jets for radio sources that are brighter towards the geometric centre, the position of the likely infrared host is expected to be in proximity to the radio emission ridge line. 
For edge-brightened sources, it should be roughly equidistant from the two lobes and lie in close proximity to the major axis of the source or ridge.
In cases where we could not locate a suitable host, we removed the source from our catalogue.

\subsubsection{Radio source characterisation}
\label{SEC:RadioClassification}
The classification of radio galaxy morphology according to \citep{fanaroff74} is typically based on two parameters: the distance between the peak emissions of the opposite lobes and the total extent of the radio emission. 
In this study, we employed a primarily manual identification process to identify these parameters, utilizing the CARTA visualization package \citep{comrie21}. 
Specifically, we measured the distance $a$ between the two emission peaks and determined the rectangular bounding box that just encompassed the 5$\sigma$ contour of the source, where $\sigma$ represents the local RMS noise. 
This bounding box is aligned with the major axis of the jet at a position angle $\theta$, and its length and width are denoted by $b$ and $c$, respectively.

We also experimented with a Laplacian of Gaussian (LoG) filter to identify the locations of the two emission peaks, determine the distance $a$ between them, measure the orientation of the major axis of the radio galaxy, and fit a bounding box as described above. However, we visually inspected each algorithm-derived fit and only considered it if it matched our manual estimate. We observed that the LoG filter produced correct results in approximately 25\% of cases.

According to the criteria of \citet{fanaroff74}, the galaxies are classified as FR-I and FR-II if $a/b < 0.5$ and $a/b > 0.5$, respectively. However, some sources have $a/b \sim 0.5$, which means a small measurement error or noise can lead to a misclassification from FR-I to FR-II or vice versa. To account for such cases, we label these sources with $a/b = 0.5\pm 0.05$ as FR-X sources, which are considered to have an unreliable classification. Consequently, FR-I and FR-II radio sources are classified with $a/b < 0.45$ and $a/b > 0.55$, respectively. Some barely resolved sources have only one peak or no peak outside the central component, and for those, $a$ is assigned as 0, and we classify them as R (for ''resolved") even though we must await future higher-resolution images to see whether they feature two jets or lobes.
Our dataset contains 328 FR-I, 128 FR-II, 110 FR-X, and 196 R radio sources, which are situated in the South-Western area of the EMU-PS survey. 
Further efforts are underway to classify about three thousand DRAGN sources in the complete EMU-PS survey \cite{yew22prep}.
Note that there are different approaches to labelling data for network training, including methods based on the number of peaks/components \citep{wu19} or the use of multiple tags for each source \citep{rudnick21}.
In the present study, we have opted to train the network using DRAGNs that are classified into FR-I, FR-II, FR-X, and R type sources following \citet{yew22prep}.
For future work, it would be valuable to compare the effectiveness of different class labels in generating segmentation masks for radio sources.

\subsubsection{Bounding Boxes and Masks for Radio Sources}
\label{SEC:BoxLabels}
The dimensions of the rectangular regions $b$, $c$, and $\theta$, are utilized to obtain bounding boxes for radio sources. 
The segmentation masks are obtained by identifying components with a signal-to-noise ratio greater than three within the bounding boxes. 
After visually examining all 762 radio sources, it was observed that in 99\% of cases, the masks within the bounding boxes correctly included all linked radio source components. 
The inaccurate masks for unrelated sources, such as nearby point sources located within the same bounding boxes, were subsequently removed from the ground truth labels.

\section{Method}
\label{SEC:method}
This section outlines the initial and important step in constructing a deep learning model: data pre-processing to make it machine-compatible. 
Additionally, we discuss the machine learning technique used in this study and model training.

\begin{figure*}
\centering
\vspace*{-0.5cm}
\includegraphics[width=15cm, scale=0.5]{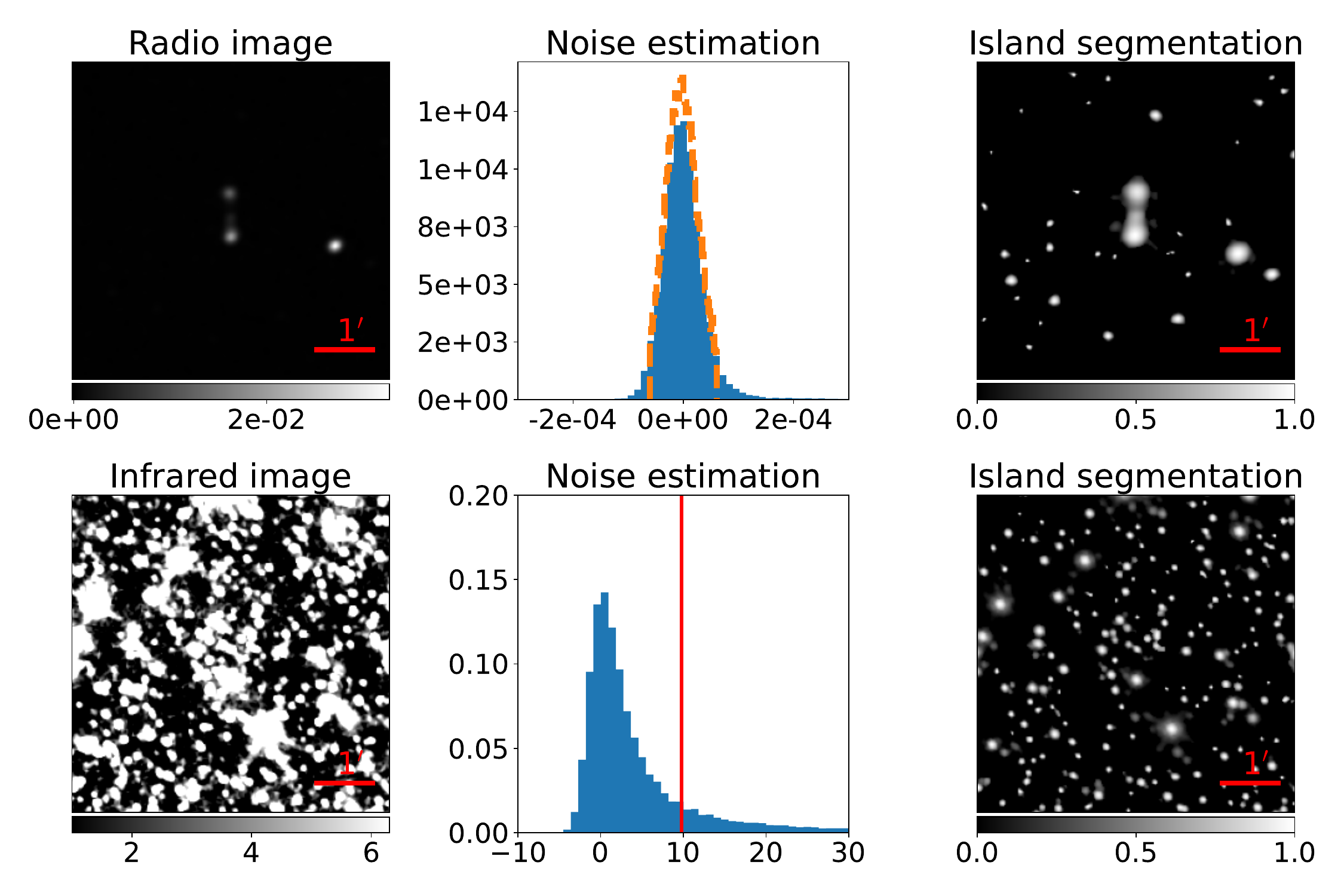}
\caption{The pre-processing methodology for radio (upper rows) and infrared (lower rows) images. Starting from the top left and moving rightwards, the first panel displays a radio image captured by ASKAP. The second panel illustrates the complete distribution of image pixels (represented by the blue-filled histogram), as well as the clipped distribution (indicated by the orange-dashed line), which is utilized to estimate the noise level as the standard deviation ($\sigma$). The third panel presents the segmented islands located where pixel values exceed 3$\sigma$. The pixel values are presented in a logarithmic scale and subjected to Min-Max normalization.  
The pre-processing steps for the infrared images, displayed in the lower rows, are identical, except that the islands are identified where the pixel value exceeds Median+3$\times$MAD, depicted by a red vertical line in the second panel.} 
\label{FIG:Preprocessing}
\end{figure*}

\subsection{Radio and Infrared Image Pre-processing}
\label{SEC:preprocessing}
The quality of data used for training machine learning models is crucial for their performance. 
The ability of a machine to identify important features and extract meaningful insights from the data is key. 
ASKAP surveys, with their high sensitivity, present unique challenges in terms of data preprocessing due to the high density of sources in the survey images. 
To address this, we have developed a pre-processing scheme that aims to enhance the relevant features in the radio images.

\begin{itemize}

\item First step is to estimate the noise in each cutout. 
To achieve this, we measure the Median Absolute Deviation (MAD) of the pixel values. 
Next, we apply two rounds of data clipping to eliminate real sources. The threshold for real sources is determined to be $3\times$MAD. After clipping, we estimate the noise as the standard deviation of the remaining distribution. 
In the top second panel of Figure~\ref{FIG:Preprocessing}, we display the full and clipped distributions of image pixels in blue (filled) and orange (dashed) colours, respectively.

\item We proceed by applying segmentation masks to the islands with pixel values greater than 3$\sigma$ in each cutout. 
Here, $\sigma$ is defined as the standard deviation of the clipped distribution. 
At the locations of these islands, we apply a logarithmic scaling to the pixel values and perform Min-Max normalization, which enhances the signal of the islands. 
The remaining pixel values in the image are set to zero, and the Min-Max normalization of the segmented regions rescales the image to the range of 0 to 1. 
In the resulting image, shown in the top third panel of Figure~\ref{FIG:Preprocessing}, the source density is moderately high. Some of the islands may be spurious noise fluctuations or artifacts.

\item To address this issue, we establish a criterion for identifying islands in the image by setting a threshold on the number of pixels they comprise. 
Specifically, we retain only those islands where the signal is dispersed across a substantial number of pixels. 
After several rounds of experimentation and visual examination, we determined that the minimum island size should be 5 pixels ($\sim10^{\prime\prime}$), eliminating most map noise fluctuations. 
However, it is essential to note that this limit may also eliminate certain low-brightness point sources. 
Nonetheless, this does not affect our analysis, as the primary aim of this study is to identify components of extended radio sources. 
\end{itemize}

WISE images underwent a similar pre-processing sequence, but the correlation among neighbouring pixels in WISE images differs from that in radio-interferometric images. 
Additionally, the WISE instrument's noise characteristics are not entirely Gaussian, and since the images are approaching the source confusion limit, the lower limit of the estimated noise is determined as Median+3$\times$MAD.
The criterion of a minimum island size of 5 pixels is also applied when identifying islands in infrared images.
After pre-processing radio and infrared images, we combine them into the 3-channel RGB images.
To do this, we first compress the original 32-bit radio and infrared images to 16-bit and 8-bit, respectively. 
Next, we fill the 8-16 bit and 0-8 bit radio information to the B and G channels, respectively, while the 8-16 bit infrared information is inserted into the R channel.

\subsection{Machine Learning Method}
\label{SEC:MLmethod}
The objective of weakly-supervised semantic segmentation (WSSS) is to reduce the cost of annotating ``strong" pixel-level masks by using ``weak" labels like bounding boxes and image-level class labels. 
This paper focuses on using image-level class labels for training, which are cost-effective but challenging. 
The pipeline involves training a multi-label classification model, extracting class activation maps, and using all-class masks as pseudo labels to learn the instance segmentation model.
Additionally, we also discuss the detection of infrared host galaxies using keypoint estimation.

\subsubsection{Class Activation Maps}
\label{SEC:CAM}
Class Activation Map \citep[CAM;][]{zhou16ML} is a technique to get the discriminative image regions used by a convolutional neural network (CNN) to identify a specific class in the image. 
In other words, a CAM lets us see which regions in the image were relevant to a particular class.
In addition, CAM also provides further insight into the network's learning process since it provides object localization for the predicted class without requiring explicit bounding box labelling by the user.
In the initial phase of CAM, a multi-label classification model is trained using global average pooling (GAP) along with a prediction layer, typically implemented as a fully connected (FC) layer. 
The prediction loss is calculated using the binary cross-entropy (BCE) loss function.
\begin{equation}
\text{BCE} = -\frac{1}{C}\sum_{k=1}^C \left[y_k\log(p_k) + (1-y_k)\log(1-p_k)\right],
\end{equation}
where $C$ is the number of object classes, $y_k$ is the true label (either 0 or 1) of the $k$-th class, and $p_k$ is the predicted probability of the positive class calculated from the sigmoid function $\sigma_S$ as
\begin{equation}
p = \sigma_S~(\text{FC}~(\text{GAP}~(f(x)))),
\end{equation}
where $x$ is the input image and $f(x)$ represents the feature map of $x$ before the GAP.
The BCE loss penalizes the model for predicting low probabilities for positive and high probabilities for negative examples.
Once the model converges, the CAM can be calculated mathematically using the following equation:
\begin{equation}
\text{CAM}_k(x) = \alpha_{k} ~ f(x),
\end{equation}
where $\text{CAM}_k(x)$ is the class activation map for class $k$, and $\alpha_{k}$ denotes the classification weights of the FC layer for class $k$.

\begin{figure*}
\centering
\includegraphics[width=18cm, scale=0.5]{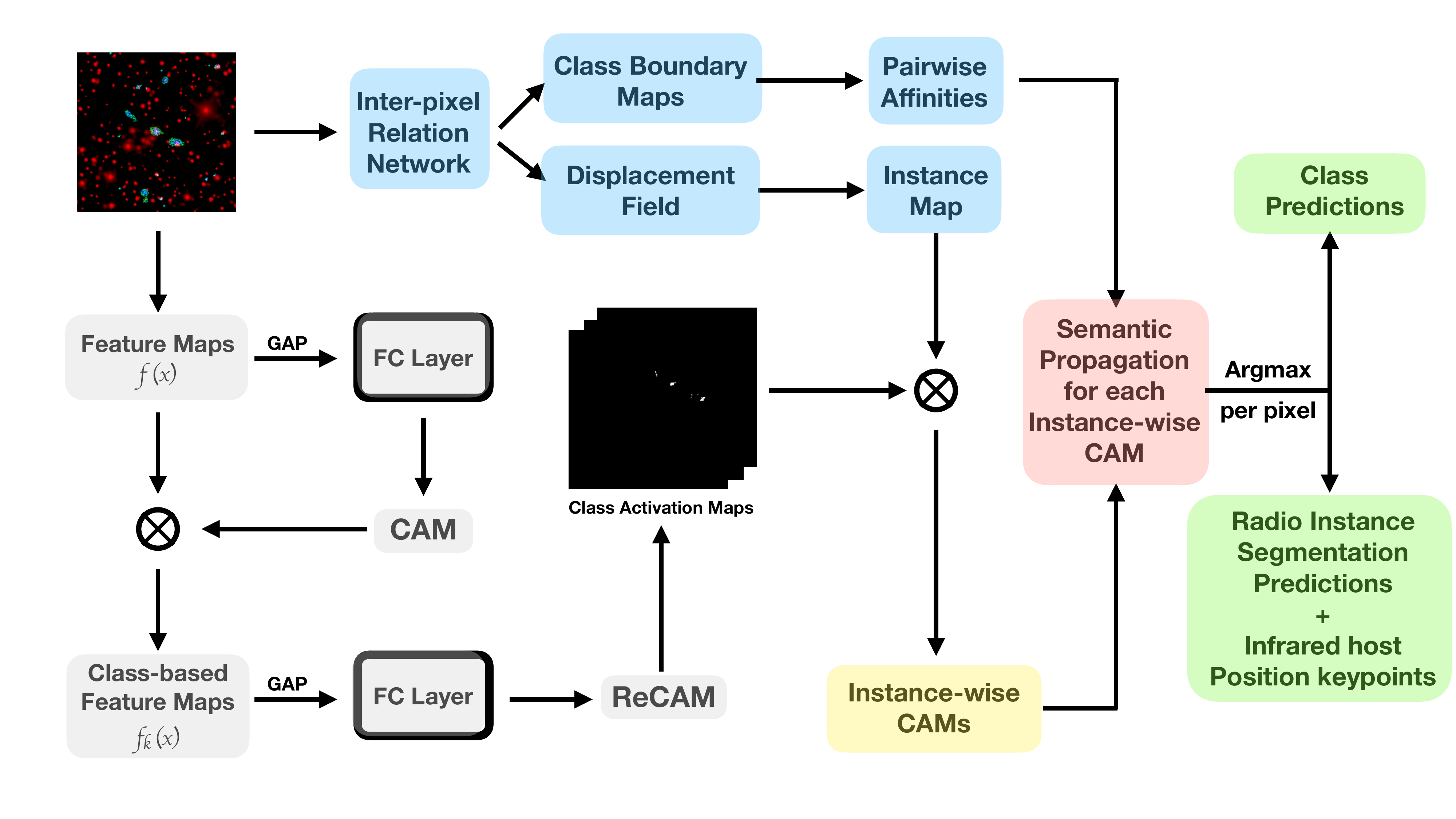}
\vspace*{-0.5cm}
\caption{Overview of the weakly-supervised framework to generate instance segmentation masks from the class labels.} 
\label{FIG:Network}
\end{figure*}

A recent study by \cite{chen22ML} has shown that the BCE may not be effective in producing satisfactory pseudo masks for the segmentation model.
Pseudo masks are artificial masks generated by applying a threshold to activation maps, with the intention of approximating the shape and position of objects in an image.
The effectiveness of BCE in generating accurate pseudo masks is hindered by the sum-over-class pooling feature, which may cause each pixel in CAM to respond to multiple classes co-occurring within the same receptive field. 
Consequently, the hot CAM pixels, which correspond to the regions in the input image that had the greatest influence on predicting a specific class, may mistakenly encompass areas that belong to other classes. 
Additionally, non-hot pixels could be inaccurately classified as belonging to the target class.

In light of this, \cite{chen22ML} proposed a solution dubbed ReCAM, which effectively reactivates the converged CAM using softmax cross-entropy loss (SCE). 
SCE involves two steps: the first step uses the softmax function to convert the model's output scores into probability distributions over the classes. 
In contrast, the second step calculates the cross-entropy between the predicted probabilities and the true class labels. 
SCE aims to minimize the difference between the predicted and actual probability distributions, thereby improving the model's accuracy in classifying the input data.
Specifically, given an image, the CAM extracts the feature pixels $f_k(x)$ for each class. 
Then it uses them and the class label to train another fully connected layer (after the backbone) using SCE. 
Once the model has converged, ReCAM is obtained in the same way as CAM. 
Since SCE is contrastive, the pixel response is divided into distinct classes, reducing ambiguity in pseudo masks.
In the present work, we use the ReCAM architecture to get the CAMs for all images containing radio sources.

\subsubsection{Inter-pixel Relation Network \& Instance Segmentation}
\label{SEC:IRN}
Instance segmentation is a computer vision task that simultaneously estimates individual objects' class labels and segmentation masks. 
While CAMs can provide a rough estimate of the areas belonging to each class by examining how local image regions contribute to the classification score of the class, they have certain limitations that prevent them from being used directly as supervision for instance segmentation. 
For example, CAMs often have limited resolution and highlight only partial areas of objects. 
Additionally, they cannot distinguish between different instances of the same class, further restricting their usefulness.
To overcome these limitations of CAMs, \cite{ahn19ML} introduced Inter-pixel Relation Network (IRNet).
The network is a weakly-supervised learning approach to instance segmentation that aims to overcome the limitations of traditional instance segmentation methods that require pixel-level annotations. Instead of relying on fully annotated datasets, this approach uses weak supervision in the form of class labels to learn to segment objects in an image.

IRNet consists of two separate branches that work together to estimate instance segmentation. 
The first branch predicts a displacement vector field, where a two-dimensional vector is assigned to each pixel in the image. 
This vector indicates the location of the centroid of the instance that the pixel belongs to. 
Using this displacement field, an instance map is generated by assigning the same instance label to pixels with displacement vectors pointing to the same location.
The second branch of IRNet detects class boundaries between different object classes by analysing the features extracted from the input image. 
It employs a boundary prediction module that takes the features extracted from intermediate layers of the network as input and focuses on capturing the distinctive patterns and characteristics that signify the boundaries between different object classes. 
By analysing these features, the boundary prediction module identifies regions in the image where transitions between object classes are likely to occur. 
To enhance accuracy, post-processing techniques like non-maximum suppression and thresholding refine the boundary map.
Using the detected boundaries, pairwise semantic affinities are computed so that two pixels separated by a strong boundary are considered a pair with a low semantic affinity. 
This allows IRNet to accurately distinguish between different object instances and produce high-quality instance segmentation results.

Subsequently, instance-wise CAMs are generated by combining CAMs with instance maps. 
These instance-wise CAMs are further improved by propagating their attention scores to related areas according to the semantic affinities between neighbouring pixels. 
Finally, an instance segmentation mask is produced by choosing the instance label with the highest attention score in the instance-wise CAMs at each pixel.

In the present work, IRNet is trained efficiently using inter-pixel relations obtained from ReCAM. This involves gathering pixels with high attention scores and using their displacements and class equivalence to train the network. 
This means that no additional supervision, besides image-level class labels, is used to train ReCAM and IRNet effectively.
Figure~\ref{FIG:Network} shows an overview of this weakly-supervised framework to generate island segmentation masks from class labels of extended radio galaxies.

\subsubsection{Keypoint Detection for Infrared Hosts}
\label{SEC:keypoints}
In addition to the instance segmentation masks for radio galaxies, we determine the most likely host galaxy from the infrared images.
We do this without relying on any supervised infrared signals. 
Our approach involves using instance-wise CAMs to initially detect all instances in the radio channel.
As the instance-wise CAMs indicate attention for radio sources, we multiply them by the corresponding pre-processed infrared channel to detect host galaxies.
At the positions of each instance mask, we find pixels where the activation maximizes in the infrared-weighted instance-wise CAMs.
These pixels are then identified as the infrared host galaxies.

\subsection{Training}
\label{SEC:Training}
The radio and infrared dataset used in this work (see Section~\ref{SEC:Observations}) is partitioned into two sets: the train and test sets. 
The train and test sets comprise 610 (80\%) and 152 (20\%) 3-channel images, along with their corresponding labels. 
The training set consists of 100 FR-I, 261 FR-II, 90 FR-X, and 159 R sources.
In the test set, there are 28 FR-I, 67 FR-II, 20 FR-X, and 37 R sources.
The labels and dataset information are compiled in a `JSON' file that contains four annotations. 
For each source, the radio annotation is stored as `categories', and `segmentation' for radio galaxy classes and segmentation masks encapsulating the full extent of radio emission, respectively. 
The positions of the infrared hosts are stored as `keypoints' that are important landmarks or features that can be used to identify specific points of interest within an image. 

The weakly-supervised deep learning model explained in Section~\ref{SEC:MLmethod} and depicted in Figure~\ref{FIG:Network}, receives 3-channel images (consisting of two radio channels and one infrared channel) and class labels as input. 
It then produces both instance segmentation masks for radio emission and positions of infrared host galaxies as output.
Our deep learning framework employs ReCAM in combination with IRNet, which utilizes class labels and can segment and locate radio galaxies and their hosts without the need for supervision.

The CAMs are generated using a combination of CAM and ReCAM networks.
The CAM network is trained first, and the output activation maps from the network are used to train ReCAM further.
We use ResNet-50 as a backbone network for both CAM and ReCAM.
ResNet-50 is a convolutional neural network architecture that addresses the problem of vanishing gradients in very deep neural networks. It has 50 layers and uses skip connections or residual connections to allow gradients to flow more easily during training \citep{he15d}. 
Using a batch size of 4 for CAM and ReCAM networks, we pass the images and class labels from the training set through these networks.
The neural network's weights are updated using gradient descent and backpropagation in batches. 
Gradient descent enables the model to minimize the difference between predicted and actual output by adjusting the model's parameters in the direction of the negative gradient of the loss function. 
Backpropagation propagates the error backward through the network to compute gradients of the loss function with respect to the model parameters. 
We use a Stochastic Gradient Descent \citep[SGD;][]{robbins1951stochastic} optimizer with a learning rate of 0.01 for the CAM network and 0.0001 for the ReCAM network to optimise the network.
The learning rate determines the step size at which the gradient descent algorithm updates the model parameters during training.
The CAM and ReCAM networks are trained for 30 and 15 epochs, respectively, where one epoch is defined as passing the entire training data once through the neural network.
As these networks are used for obtaining attention maps, they require a small number of epochs for training. 
This also helps to reduce the risk of over-fitting.

Moreover, to improve the generalization capability of the CAM and ReCAM networks, prevent overfitting, and further enhance the dataset's balance, we incorporate data augmentations during the training phase. 
We incorporate a random set of augmentations at each iteration of the training phase to expose the networks to a diverse range of data representations and scenarios. These augmentations include resizing the images, which helps to standardize their dimensions. By resizing the images, we bring them to a common scale, allowing the network to learn and extract features irrespective of variations in size. Another important augmentation is random cropping, which involves selecting a random region of the image. By randomly cropping the images around the centre, we provide the network with a rich set of training examples that encompass various placements and arrangements of the radio sources. Furthermore, random flipping and rotation of the images are employed to expose the network to patterns and features in various orientations. By incorporating these diverse augmentations, the network becomes more adept at handling a wide range of image variations. It learns to generalize well to different scales, positions, and orientations of radio sources, thus enhancing its overall predictive power and adaptability to different scenarios.

\begin{figure}
\centering
\includegraphics[trim=0cm 0cm 0cm 1cm, width=4.2cm, scale=0.6]{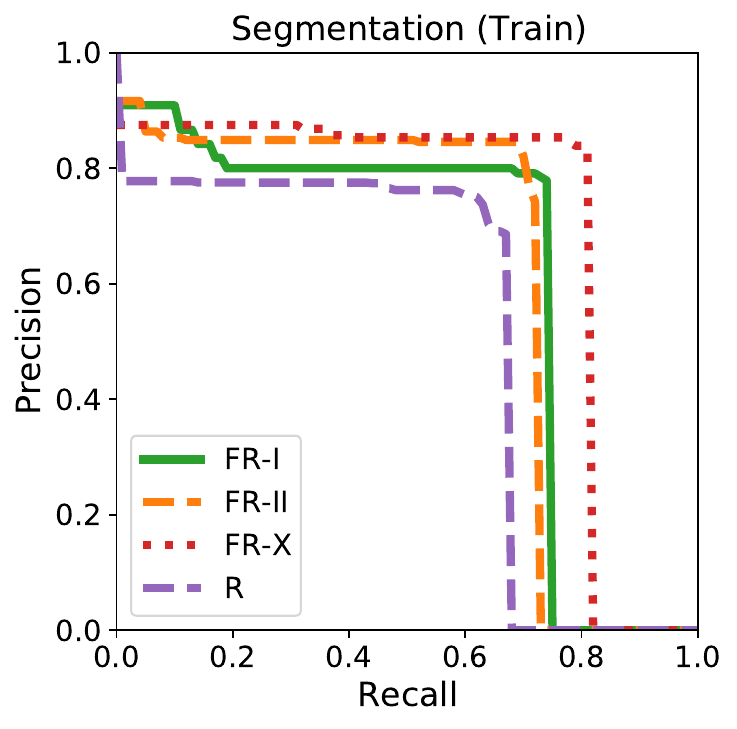}
\includegraphics[trim=0cm 0cm 0cm -1cm, width=4.2cm, scale=0.6]{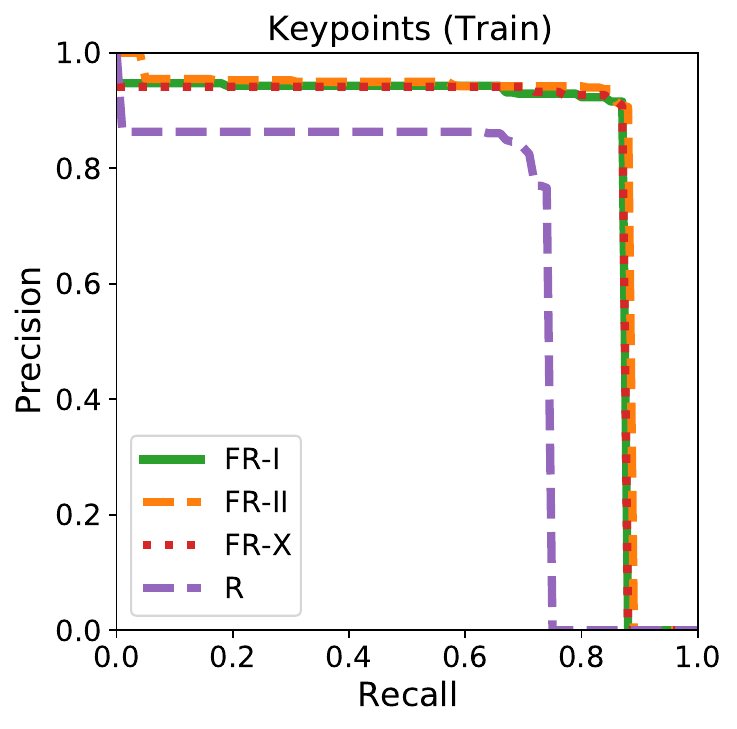}
\includegraphics[trim=0cm 0cm 0cm 1cm, width=4.2cm, scale=0.6]{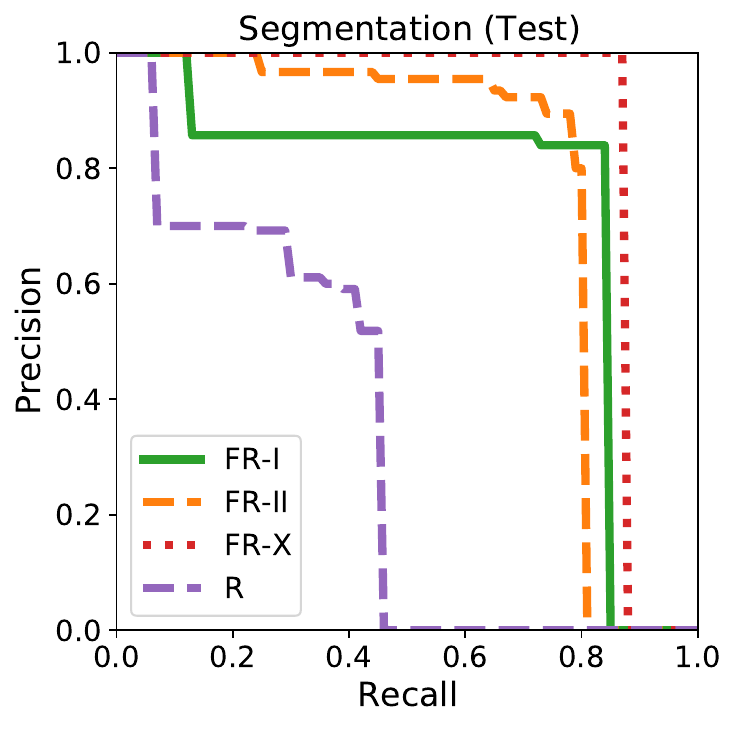}
\includegraphics[trim=0cm 0cm 0cm -1cm, width=4.2cm, scale=0.6]{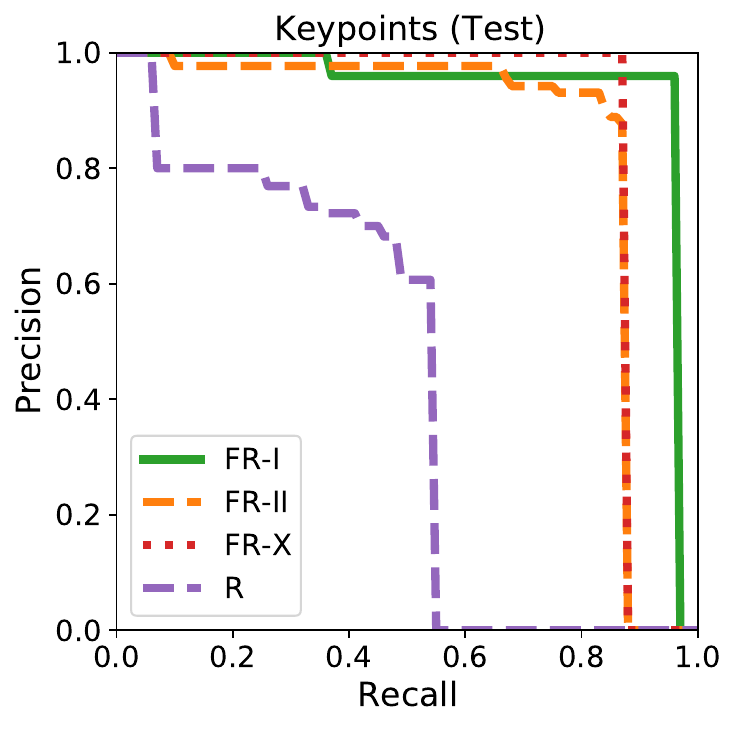}
\caption{The figure illustrates precision-recall curves for segmentation (first column) and keypoint detection (second column) tasks, covering both the training and testing samples. Each curve represents the precision-recall values for the four classes. These curves provide a comprehensive view of the model's performance across different classes and tasks, demonstrating its precision and recall capabilities.} 
\label{FIG:PRcurves}
\end{figure}

The IRNet architecture has two output branches: one predicts a displacement vector field, and the other predicts a class boundary map.
Both branches use the same ResNet-50 backbone and take feature maps from the backbone as inputs. 
The convolution layers in both branches are followed by group normalization \citep{wu2018group} and ReLU \citep{agarap2018deep}, except for the last layer. 
The displacement field branch uses a top-down pathway to merge the feature maps, upsample the low-resolution feature maps, concatenate them, and process them with a $1\times1$ convolution layer. 
Finally, a displacement field is decoded through three $1\times1$ convolution layers. 
The boundary detection branch uses a $1\times1$ convolution layer for dimensionality reduction, and the results are concatenated and fed into the last $1\times1$ convolution layer, which produces a class boundary map.
We use stochastic gradient descent to optimise the network and apply polynomial decay \citep{liu2015parsenet} to decrease the learning rate at each iteration. 
The network is trained with a batch size of 16 for three epochs.
We start with a learning rate of 0.1. During training, the ResNet-50 backbone of IRNet remains frozen, and the displacement field branch receives gradients that are amplified by a factor of 10.
Training for the CAM, ReCAM, and IRNet models is performed on a cluster with two NVIDIA P100 GPUs and 16 GB of memory and takes approximately two hours to complete.

\section{Results}
\label{SEC:Results}
Our weakly-supervised deep learning model, trained using images and weak image-level labels, predicts pixel-level information in the radio and infrared channels.
The weak image-level labels used for training are the radio source classes, i.e. FR-I, FR-II, FR-X and R radio sources (see Section~\ref{SEC:Labels}).
The expected predicted pixel-level information includes masks for the extended radio emission encapsulating all galaxy components and the positions of the infrared host galaxies.
Our objective is to assess the effectiveness of a weakly-supervised deep learning algorithm with limited labelled data.
Thus, we exclusively use class-level labels during training and utilize pixel-level labels solely for model inference.
Therefore, we report the results for both training and test datasets as the networks are trained on class-level labels only and the predictions are at the pixel level.
While we do not expect weakly-supervised networks to perform as well as supervised networks due to the lack of training supervision for pixel-level labels, we observe that our weakly-supervised network achieves high accuracy in predicting pixel-level information.

\begin{table}[]
 \centering
  \begin{NiceTabular}{crlllll}
   \hline
    \hline
    &                      & FR-I  & FR-II & FR-X  & R & All \\
    &                      & AP$_{50}$  & AP$_{50}$ & AP$_{50}$ & AP$_{50}$ & mAP$_{50}$ \\
    &                      & (\%)  & (\%)  & (\%)  & (\%) & (\%) \\
    \midrule
    \Block{2-1}{\rotate Train}
    & Segmentation         & 60.9  & 61.5  & 70.0  & 51.8  &  61.1 \\
    & Keypoints            & 81.9  & 83.7  & 81.8  & 63.8  &  77.8 \\ 
    \midrule
    \Block{2-1}{\rotate Test}
    & Segmentation         & 73.8  & 77.1  & 87.1  & 32.0  &  67.5 \\
    & Keypoints            & 93.7  & 84.3  & 87.1  & 42.2  &  76.8 \\
    \hline
    \end{NiceTabular}
    \caption{Comparison of the IRNet predicted radio segmentation masks and infrared host position keypoints with the ground truth for training and test datasets.
    The AP$_{50}$ refers to the average precision for each class at IoU and OKS thresholds of 0.5. 
    mAP$_{50}$ is the mean AP$_{50}$ over all classes.}
    \label{TAB:AP}
\end{table}

\begin{figure*}
\centering
\vspace*{-1.5cm}
\includegraphics[trim=4cm 8cm 4cm 1cm, width=18cm, scale=0.5]{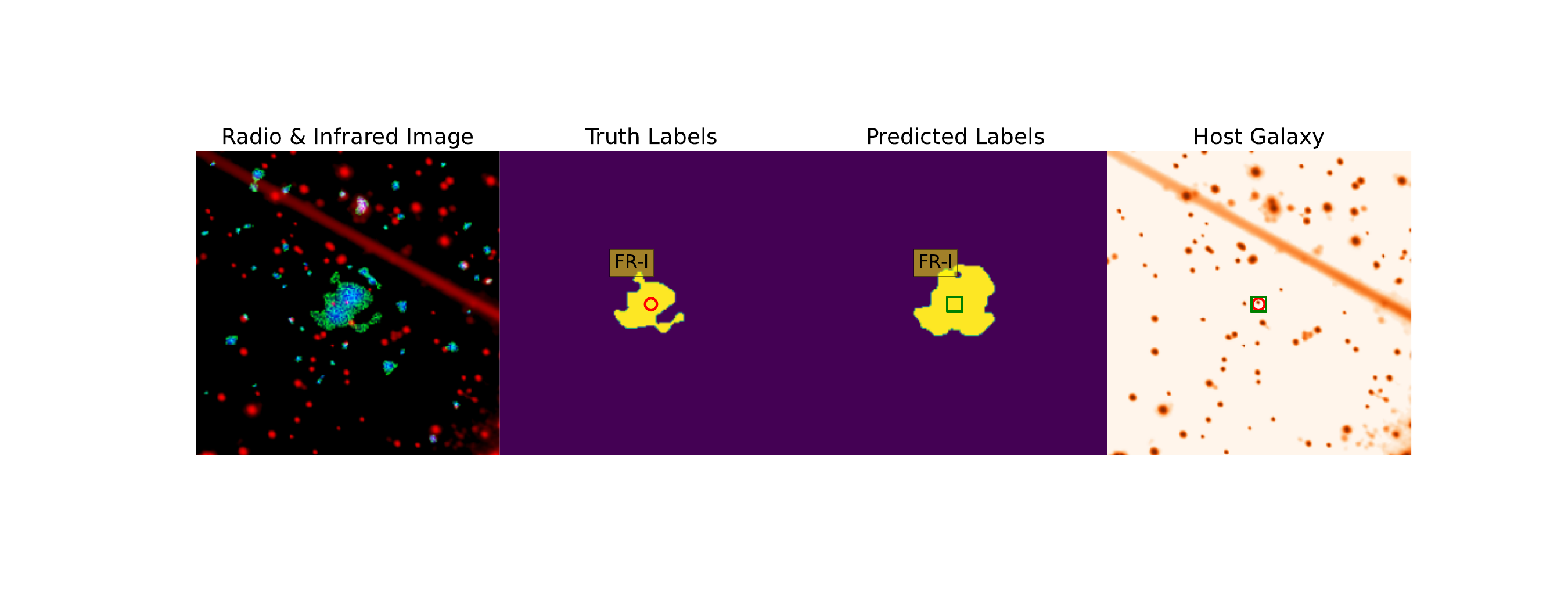}
\includegraphics[trim=4cm 8cm 4cm 0.9cm, width=18cm, scale=0.5]{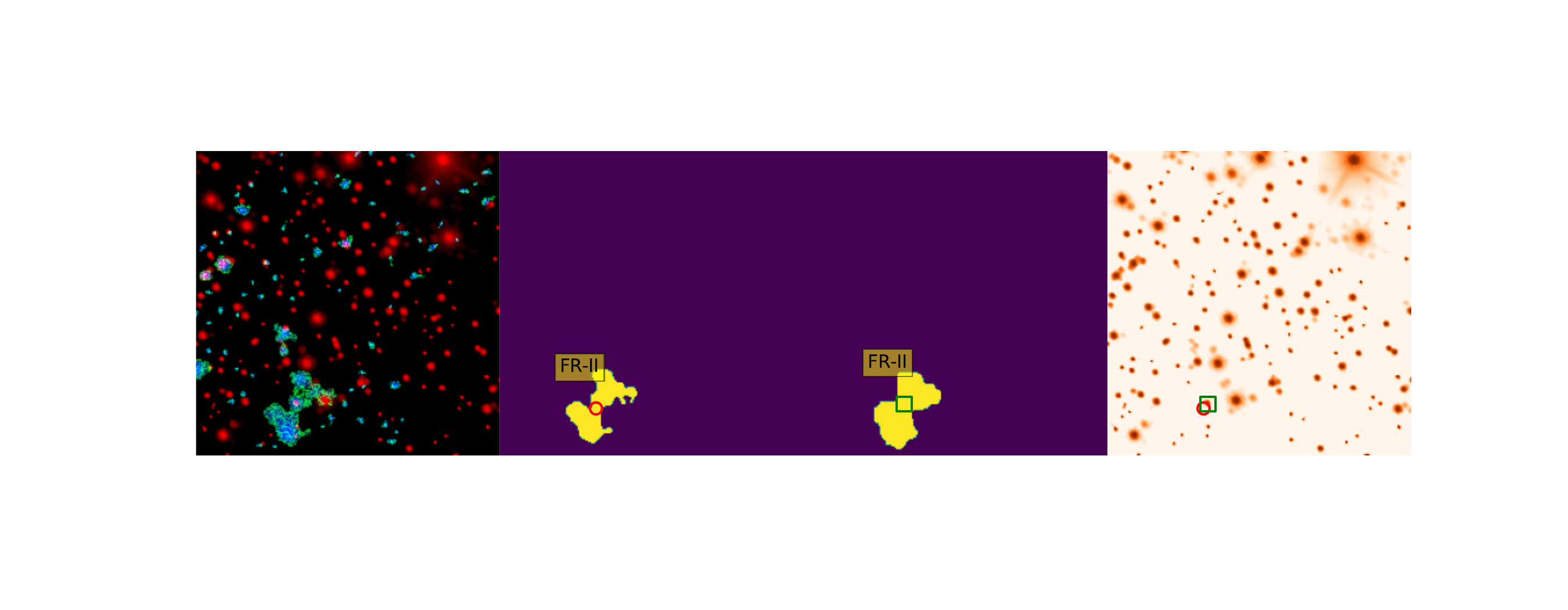}
\includegraphics[trim=4cm 8cm 4cm 0.9cm, width=18cm, scale=0.5]{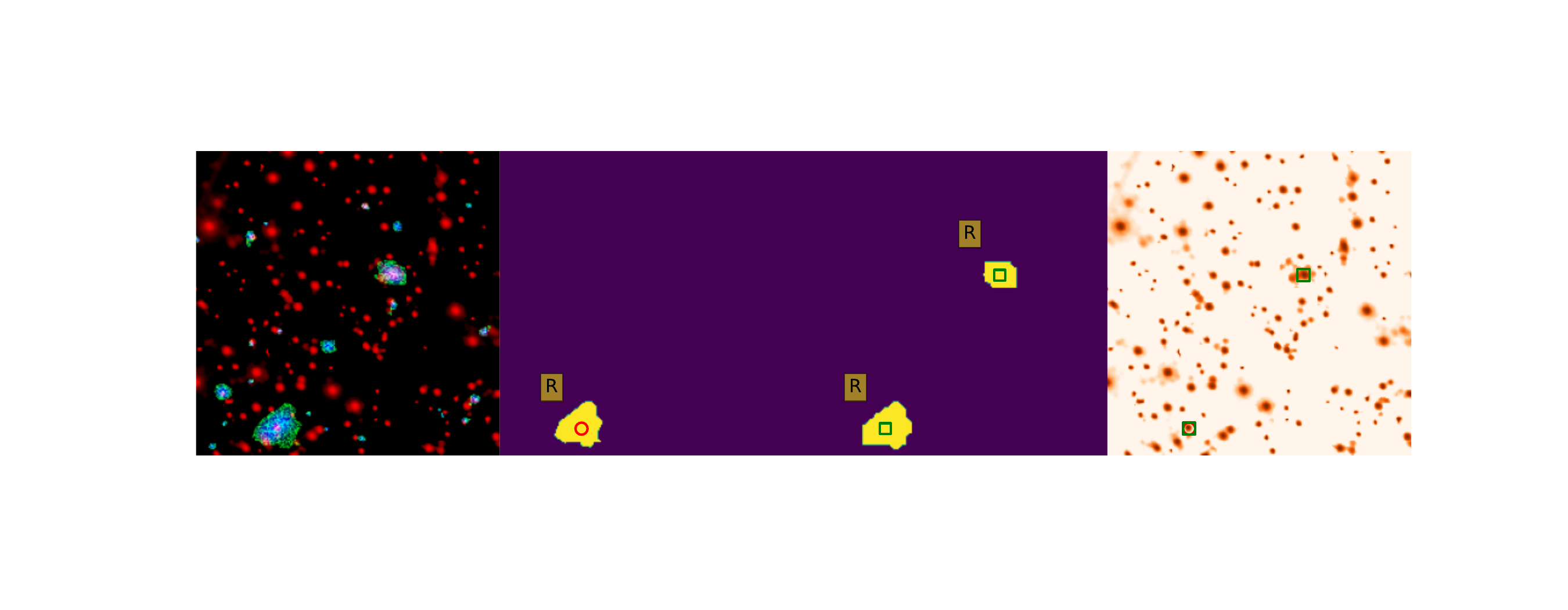}
\includegraphics[trim=4cm 8cm 4cm 0.9cm, width=18cm, scale=0.5]{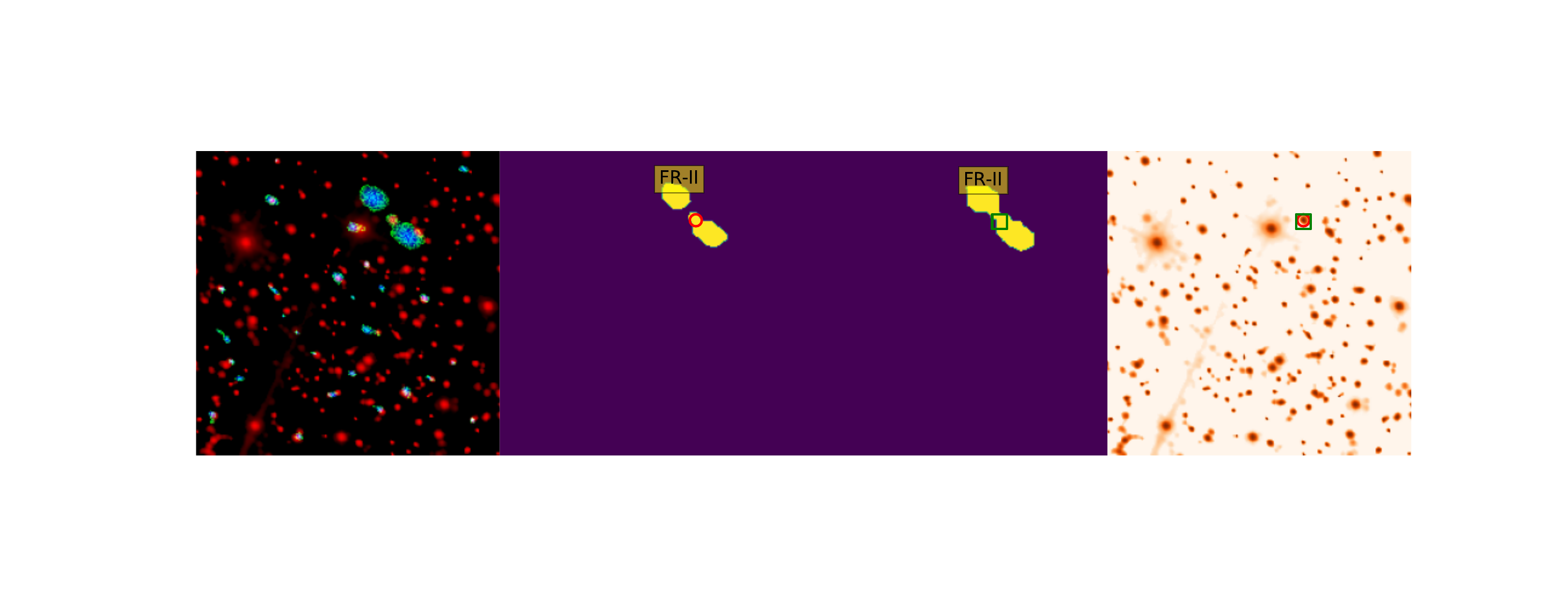}
\includegraphics[trim=3.9cm 0cm 4.1cm 0.9cm, width=18cm, scale=0.5]{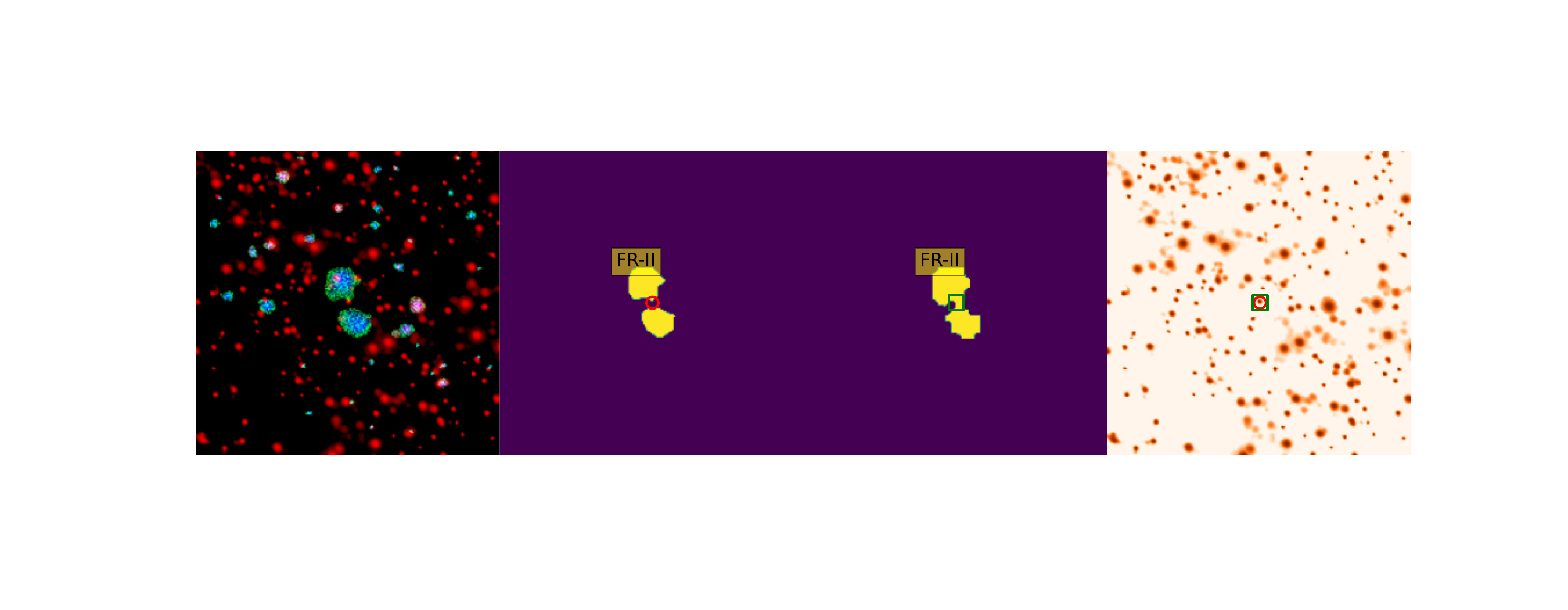}
\vspace{-2cm}
\caption{Shown in the rows from top to bottom are examples of FR-I, FR-II and R radio sources, along with the precisely predicted labels obtained using our weakly-supervised network.
Each row includes a 3-channel image of the corresponding radio (blue-green) and infrared (red) sky region in the first column, truth labels for the radio galaxy classes, segmentation masks over the radio emissions (yellow), and infrared host galaxy positions (pink circles) in the second column (see Section~\ref{SEC:Labels}).
The third column displays predicted segmentation masks (yellow) and infrared hosts (green squares), while the fourth column only shows the predicted positions of infrared hosts (green squares) and ground truth positions (pink circles) overlaid on the corresponding infrared images.
It is worth noting that our network is trained solely with class labels, yet it is capable of predicting both the radio masks and infrared host positions.} 
\label{FIG:Results1}
\end{figure*}
\begin{figure*}
\centering
\vspace*{-1.5cm}
\includegraphics[trim=4cm 8cm 4cm 1cm, width=18cm, scale=0.5]{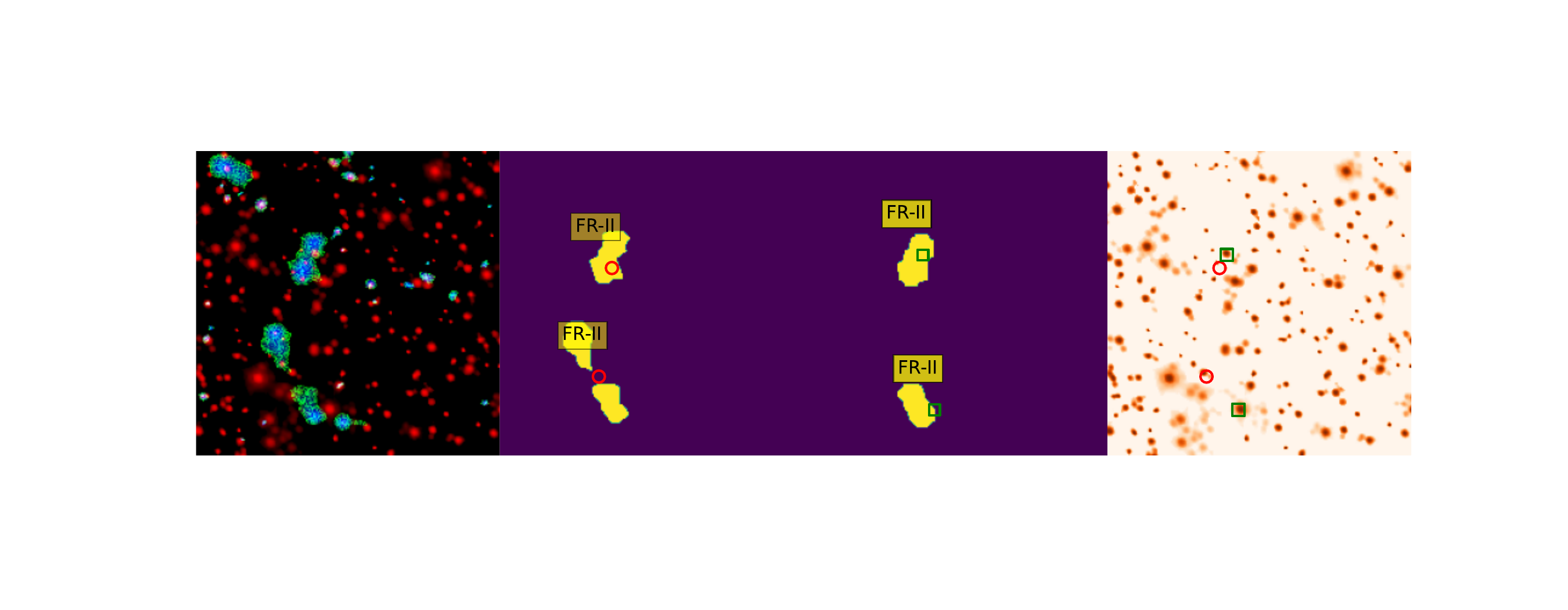}
\includegraphics[trim=4cm 8cm 4cm 0.9cm, width=18cm, scale=0.5]{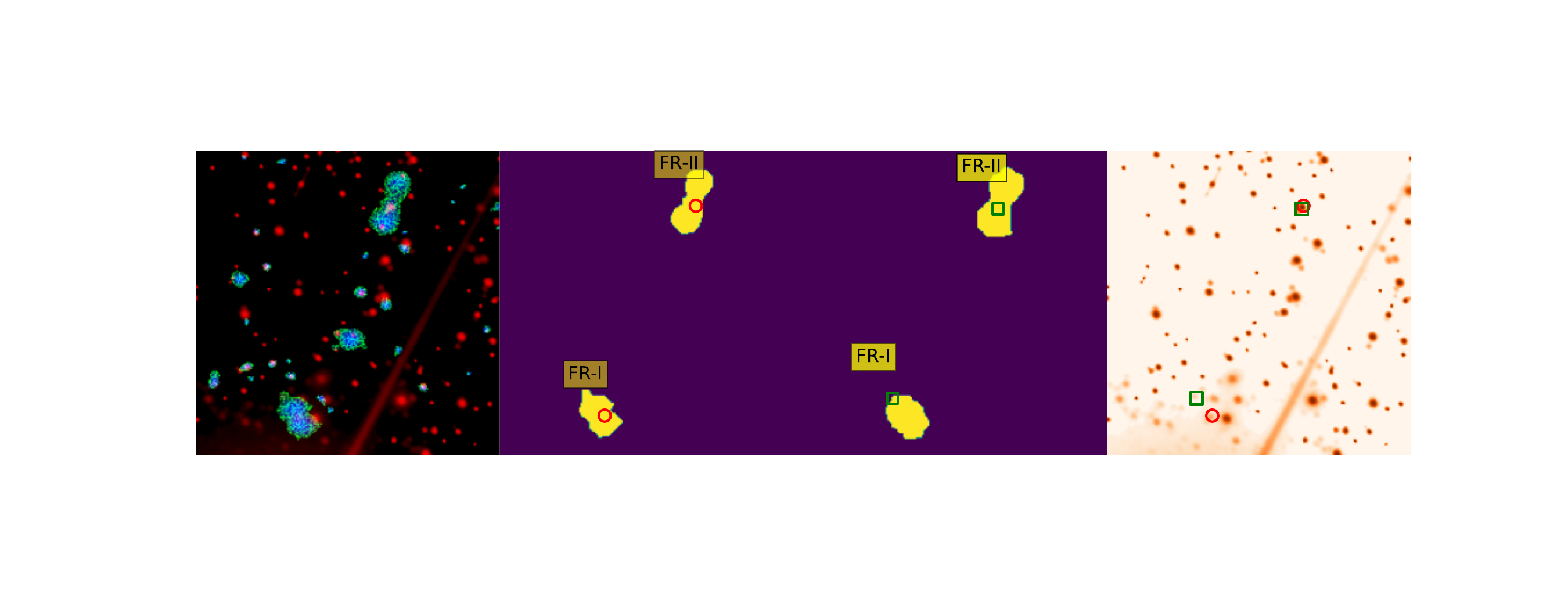}
\includegraphics[trim=4cm 8cm 4cm 0.9cm, width=18cm, scale=0.5]{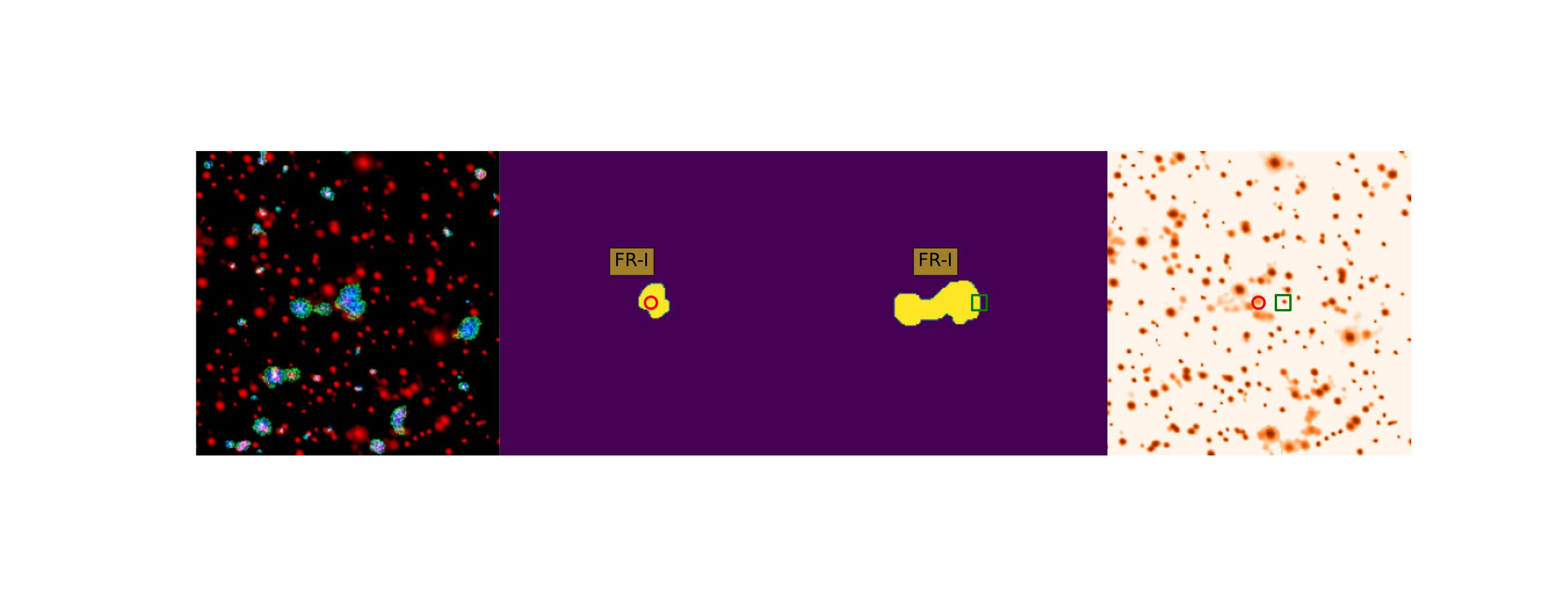}
\includegraphics[trim=4cm 8cm 4cm 0.9cm, width=18cm, scale=0.5]{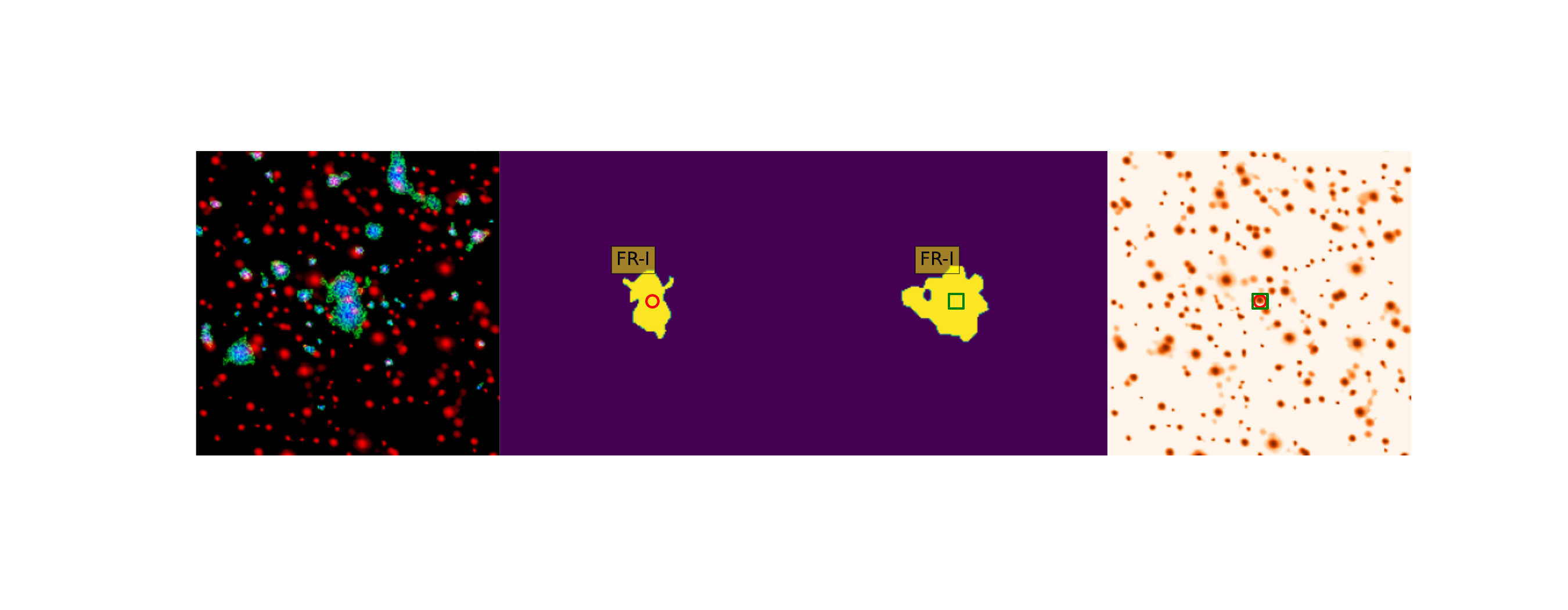}
\vspace{-2cm}
\includegraphics[trim=4cm 0cm 4cm 0.9cm, width=18cm, scale=0.5]{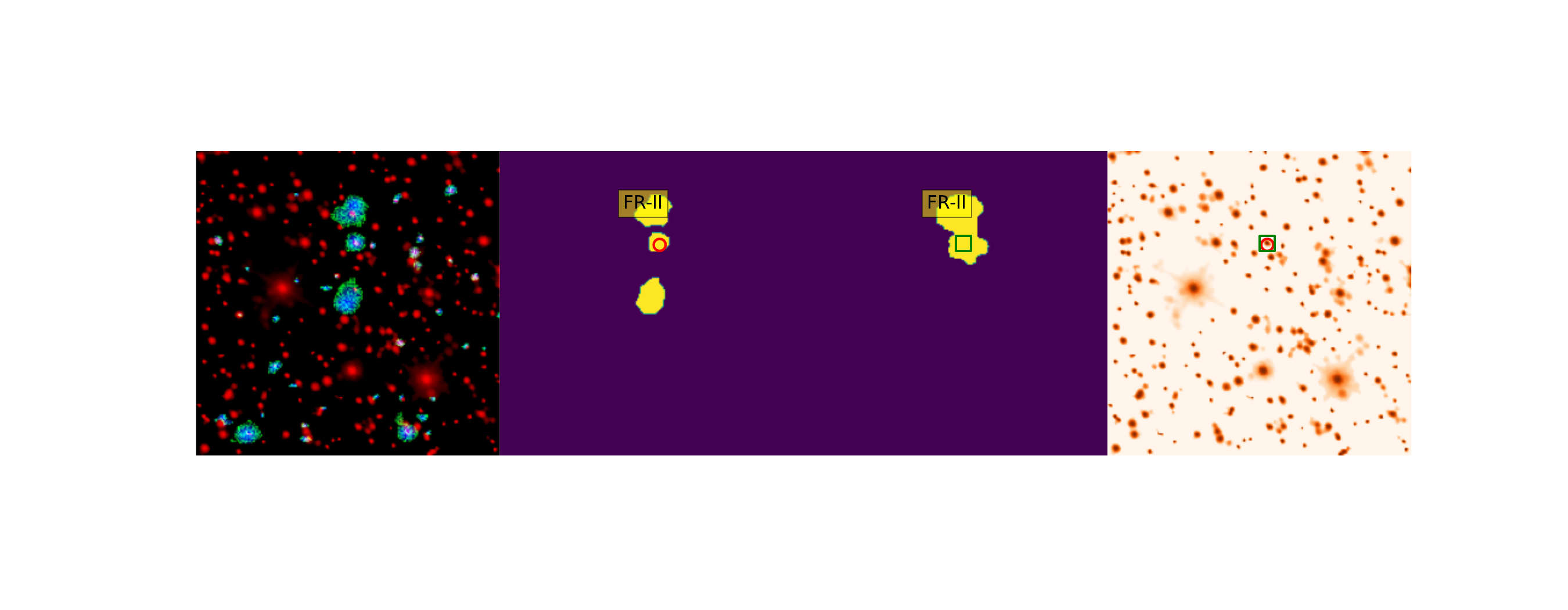}
\caption{The presented examples showcase FR-I and FR-II radio sources where the network fails to predict labels precisely. 
The columns align with those in Figure~\ref{FIG:Results1}.
The top two rows show multiple radio sources per image where the network's predictions are inaccurate. In the first row, the network fails to predict one of the lobes of an FR-II source. In both rows, the network also struggles to accurately predict the infrared hosts because of a brighter nearby infrared galaxy.
In the third and fourth rows, the network sometimes fails to predict the correct radio masks in the case of adjacent double-lobed or point sources. 
Additionally, for FR-II type sources, the network sometimes cannot predict a complete radio mask for all components when a component is too far from the central host emission, as shown in the lowest row.
} 
\label{FIG:Results2}
\end{figure*}

We use the benchmark evaluation metrics defined by \cite{coco14} for computer vision tasks.
We estimate the Intersection over Union (IoU), a metric used to evaluate the performance of image segmentation algorithms. 
The mean Intersection over Union (mIoU) is calculated as the mean of the intersection over union values for each class. 
The intersection over union is the ratio of the intersection between the predicted and true segmentation masks to the union of the two masks.
The metric can be written as
\begin{equation}
    \text{mIoU} = \frac{1}{nC}\sum_{c \in C} \sum_{i=1}^{n} \frac{{TP}_c^{(i)}}{{TP}_c^{(i)} + {FP}_c^{(i)} + {FN}_c^{(i)}},
\end{equation}
where $C$ represents all the classes in the dataset, $n$ is the total number of samples in the dataset, and $TP_c^{(i)}$, $FP_c^{(i)}$, and $FN_c^{(i)}$ represent the number of true positive, false positive, and false negative predictions, respectively, for class $c$ on sample $i$.
In the context of image segmentation, each pixel in the predicted segmentation is evaluated to determine whether it is a true positive (TP), false positive (FP), or false negative (FN) with respect to the ground truth segmentation. 
Specifically, a pixel is considered TP when it is correctly predicted as part of the object in the ground truth, FP when it is predicted as part of the object but is actually not part of it in the ground truth, and FN when it is part of the object in the ground truth but not predicted as such by the model.
We measure $\rm mIoU=40.8\%$ and 38.4\% for training and test datasets, respectively, for class activation maps generated from the ReCAM network.

For keypoint detections, we estimate Object Keypoint Similarity (OKS) metric. 
OKS measures the similarity between predicted keypoint locations and ground-truth keypoint locations.
The OKS metric is calculated by first computing the Euclidean distance between each predicted keypoint and its corresponding ground-truth keypoint, normalized by the size of the object instance. Then, a Gaussian function is applied to each distance as
\begin{equation}
\mathrm{OKS} = \exp \left(-\frac{d^2}{2s^2k^2}\right),
\end{equation}
where $d$ is the euclidean distance between the ground truth and predicted keypoint, $s$ is the area of the bounding box divided by the image cutout area, and $k$ is keypoint constant set to 0.5. 
The resulting OKS score is between 0 and 1, where 1 indicates perfect keypoint localization.

We use Average Precision (AP) to measure the performance of instance segmentation radio masks and the infrared host positions generated by IRNet.
AP is a standard metric used to evaluate object detection or segmentation models.
Precision is the ratio of true positives (correctly identified objects) to the total number of objects identified as positive (both true positives and false positives). 
The recall is the ratio of true positives to the total number of objects with truth labels. The precision-recall curve is a graph that shows the trade-off between precision and recall as the detection threshold is varied.
Figure~\ref{FIG:PRcurves} illustrates the precision-recall curves for every class, covering both segmentation and keypoint tasks. The curves represent the performance on both the training and testing samples, providing a comprehensive evaluation of the model's precision and recall capabilities.
The precision-recall curve is first computed for a set of test images to calculate AP. 
Then, the area under the curve (AUC) is calculated, averaging this value across all classes or objects of interest. 
AP is in a range of 0 and 1, and a higher AP indicates better model performance regarding precision and recall.
We calculate mean average precision (mAP), the average of AP values across multiple classes at a standard IoU and OKS threshold of 0.5 (mAP$_{50}$) for segmentation masks and keypoints, respectively.
The IoU threshold describes that the prediction is correct if its intersection over union with the ground truth is above this value.
The OKS threshold means that a predicted keypoint is considered correct if its similarity score is above this value.

The instance segmentation and positional accuracy of infrared host keypoints for the training set achieve a mAP$_{50}$ of 61.1\% and 77.8\%, respectively. Similarly, for the test dataset, we obtain a mAP${50}$ of 67.5\% and 76.8\% for instance segmentation and positional accuracy of infrared host keypoints, respectively.
Tabel~\ref{TAB:AP} also shows the AP$_{50}$ values for each radio galaxy class.
It is worth noting that the AP$_{50}$ values for the R sources are lower than those of other sources. 
While the exact reason behind this is difficult to determine, one possibility is the unique small-scale characteristics of R sources, such as having one peak or no clear peak outside the central component. 
As a result, augmentations like random flipping and rotation may have a smaller impact on these sources compared to the FR-I, FR-II and FR-X sources, which have two distinct peaks.

Examples of galaxies with precisely predicted radio masks and infrared hosts by our weakly-supervised network are demonstrated in Figure~\ref{FIG:Results1}.
The presented examples show the predicted labels of radio sources obtained through our weakly-supervised network.
Each row depicts a 3-channel image of the radio (blue-green) and infrared (red) sky region, followed by the truth labels for radio source classes, segmentation masks for radio emissions, and positions of infrared host galaxies. 
The third column displays all predicted labels. 
The fourth column shows only the predicted positions of infrared hosts (in green squares) overlaid on the corresponding infrared images, along with the ground truth positions (in pink circles).
To showcase the predictive power of our network, we examine three specific scenarios. 
In the first scenario, we deliberately position the radio sources near the edges of the image. 
Despite the network being trained on images with sources at the centre, it can accurately predict the positions and characteristics of these edge sources. 
This is made possible by the random cropping augmentations applied during training, which simulate the presence of sources away from the centre.
In the second scenario, we consider images with multiple radio sources. 
Although there are only a limited number of such images in our training and test sets (39 and 13, respectively), the network exhibits the ability to locate the presence of multiple sources within a single image.
The third scenario involves images where the radio sources are positioned at the centre.

The first row displays an FR-I radio galaxy, where the network's predicted masks encompass the entire radio region. 
The second row shows an FR-II radio galaxy near the edge of the image, where the predicted and true radio masks have the same orientation angle in the sky. 
In the third row, the network successfully detected an R source near the edge. Interestingly, it also identified an additional small-scale R source towards the centre, despite its absence in the labelled data.
The fourth and fifth rows exhibit FR-II galaxies with radio components separated by some distance.
Even though there is no visible connection between the lobes in one case and one FR-II is located near the edge in the other case, our network accurately predicts these galaxies' masks, orientation, and morphological details.
The network predicts correct infrared host galaxy positions in all these examples.

Contrary to the highly accurate predictions, Figure~\ref{FIG:Results2} shows examples where the network fails to predict the radio masks and/or infrared positions precisely.
In the top two rows, each image contains multiple radio sources. In the first row, the network fails to predict one of the lobes of an FR-II source correctly. Additionally, in both rows, the network inaccurately predicts the positions of the infrared hosts due to the presence of a brighter nearby infrared galaxy. 
In the third and fourth rows, the network fails to predict the precise radio masks in the presence of nearby double-lobed and point sources. 
Moreover, for FR-II galaxies, the network occasionally fails to generate a complete radio mask for all components when a component is far from the central host emission, as shown in the final row.

\section{Conclusions}
\label{SEC:conclusions}
Recent advancements in radio astronomy have allowed for deep continuum imaging of large areas of the radio sky, producing multi-million catalogues of radio galaxies. 
However, these surveys have also led to the detection of increasingly complex radio galaxies with multiple components, making it challenging to identify all related components using traditional methods powered by human intelligence. 
In recent years, machine learning has emerged as a powerful tool for extracting and modelling high-dimensional information from images. 
While image tags are informative and cost-effective, they cannot be used to generate pixel-level information, such as segmentation masks, crucial for identifying related radio components in images.

The present work discusses the weakly-supervised semantic segmentation (WSSS) technique that aims to reduce the cost of annotating pixel-level masks using weak image-level class labels. 
We use image-level class labels to train a multi-label classification model, which then extracts class activation maps (CAMs) to learn the instance segmentation model. 
We also discuss the Inter-pixel Relation Network (IRNet) for instance segmentation, which overcomes the limitations of traditional supervised methods that require pixel-level annotations. 
Additionally, we discuss the detection of infrared host galaxies using keypoint estimation. 

We use the data from Australian Square Kilometre Array Pathfinder (ASKAP) telescope.
The ongoing Evolutionary Map of the Universe (EMU), aims to observe the entire Southern Sky. 
The EMU Pilot Survey (EMU-PS) was conducted in 2019, covering a sky area of 270 square degrees with an RMS sensitivity of 25-35 $\mu$Jy/beam. 
In this pilot survey, a catalogue of Double Radio sources Associated with Active  Galactic Nuclei (DRAGN) was created, with source identification conducted through visual inspections \citep{yew22prep}. 
We obtain infrared images at the exact locations of the radio source cutouts using the Wide-field Infrared Survey Explorer. 
The radio and infrared image cutouts contain four types of labels, including class-level labels indicating the types of radio galaxies and pixel-level labels that include bounding boxes placed at the locations of radio galaxies, segmentation masks for radio galaxies, and positions of host galaxies in the infrared. 
The objective of the study is to evaluate the efficacy of a weakly-supervised deep learning algorithm trained with weakly labelled data.

The weakly-supervised deep learning model receives 3-channel images (RGB), which consist of two radio channels (G and B channels) and one infrared channel (R channel), and four radio source class labels as input. 
The deep learning framework employs ResNet-50 as a backbone network for both CAM and ReCAM networks, and IRNet architecture has two output branches, one predicts a displacement vector field, and the other predicts a class boundary map. 

We show that the weakly-supervised network achieves high accuracy in predicting pixel-level information, including masks for the extended radio emission encapsulating all galaxy components and the positions of the infrared host galaxies. The predicted positions of infrared hosts are demonstrated in Figure~\ref{FIG:Results1} for FR-I and FR-II radio galaxies.

We show that our weakly-supervised deep learning model predicts pixel-level information in the radio and infrared channels of galaxies using weak image-level labels (FR-I, FR-II, FR-X and R radio galaxies) for training. 
The model achieves high accuracy in predicting pixel-level information, including masks for extended radio emissions and the positions of infrared host galaxies. 
We show examples of accurately predicted radio masks and infrared hosts by the weakly-supervised network and examples where the network fails to predict these features precisely.
We use Average Precision (AP) to measure the performance of instance segmentation masks and infrared host galaxy position keypoints generated by IRNet. 
We calculate mean average precision (mAP), which is the average of AP values across multiple classes at a standard IoU threshold of 0.5 (mAP$_{50}$).
The test dataset shows mAP$_{50}$ of 67.5\% and 76.8\% for radio masks and infrared host positions, respectively.
Although these results are promising considering the weakly-supervised approach used, further research should focus on improving pixel-level detections to make the method more suitable for next-generation large-scale catalogues. 
Additionally, future studies should explore supervised and semi-supervised approaches to address the radio component association problem.
Research efforts should expand to apply machine learning techniques to other types of radio galaxies, such as those with peculiar radio morphologies, and other radio surveys with different resolutions and noise properties.
Furthermore, future studies should compare machine learning methods with traditional source finders to assess their ability to efficiently group associated components of radio galaxies and create consolidated catalogues.

\section{Acknowledgements}
The Australian SKA Pathfinder is part of the Australia Telescope National Facility, which is managed by CSIRO. The operation of ASKAP is funded by the Australian Government with support from the National Collaborative Research Infrastructure Strategy. ASKAP uses the resources of the Pawsey Supercomputing Centre. The establishment of ASKAP, the Murchison Radio-astronomy Observatory and the Pawsey Supercomputing Centre are initiatives of the Australian Government, with support from the Government of Western Australia and the Science and Industry Endowment Fund. We acknowledge the Wajarri Yamatji people as the traditional owners of the Observatory site.
The photometric redshifts for the Legacy Surveys (PRLS) catalogue used in this paper were produced thanks to funding from the U.S. Department of Energy Office of Science and Office of High Energy Physics via grant DE-SC0007914.
This research has made use of the NASA/IPAC Extragalactic Database (NED), which is operated by the Jet Propulsion Laboratory, California Institute of Technology, under contract with the National Aeronautics and Space Administration.
NG acknowledges support from CSIRO’s Machine Learning and Artificial Intelligence Future Science (MLAI FSP) Platform. HA has benefited from grant CIIC 138/2022 of Universidad de Guanajuato, Mexico.

\bibliography{ASKAP_PASA}

\appendix
\section{Confusion Matrices for Radio Source Instances}
\label{SEC:Confusion}

\begin{figure}
\centering
\includegraphics[trim=0cm 0cm 0cm 0cm, width=8cm, scale=0.5]{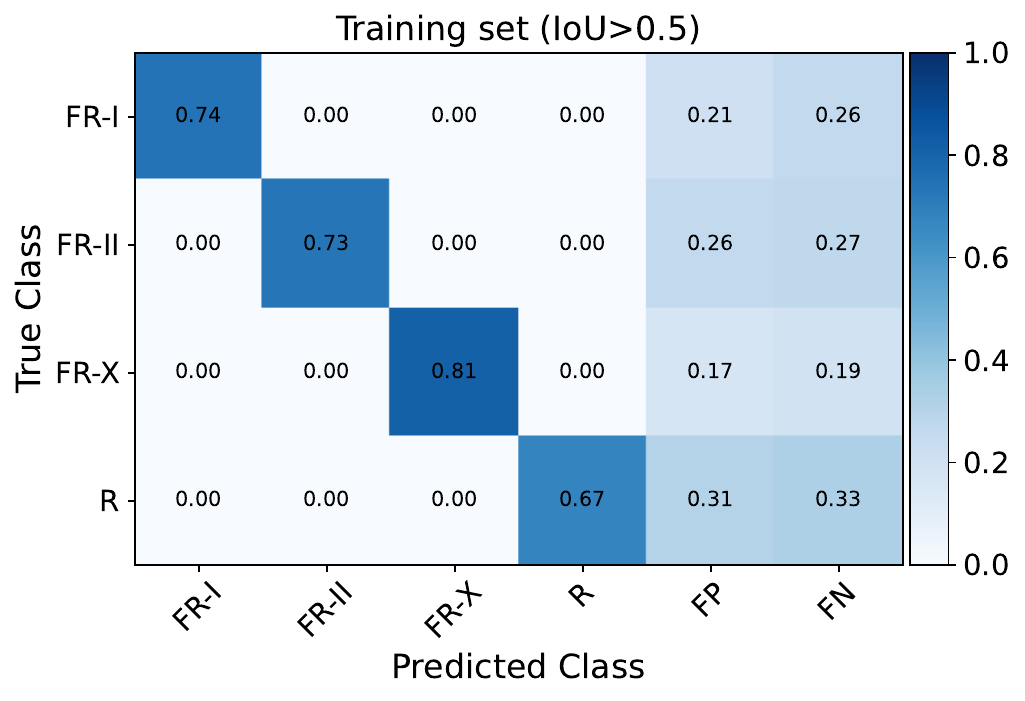}
\includegraphics[trim=0cm 0cm 0cm 0cm, width=8cm, scale=0.5]{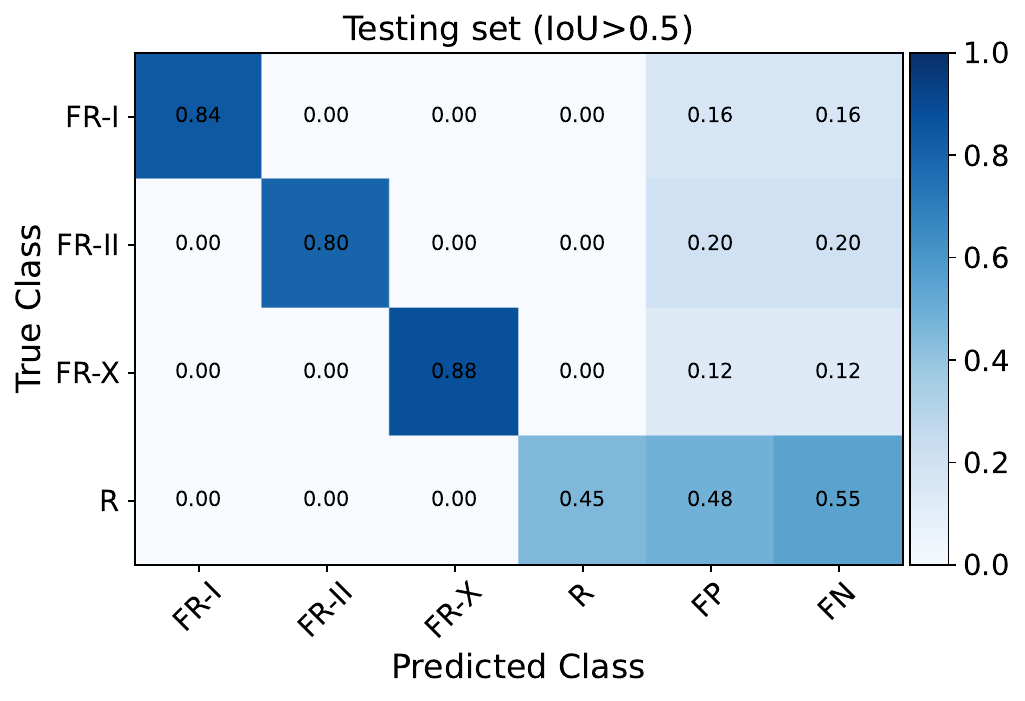}
\caption{Shown are the confusion matrices for the training (top) and testing (bottom) datasets. 
The matrices are normalized based on the total number of sources for each class. 
The diagonal values in the matrices indicate TP instances, representing objects that are correctly detected and accurately segmented with an IoU threshold above 0.5 compared to the ground truth instances.
FP instances correspond to model detections without corresponding ground truth instances, while FN instances represent objects missed or undetected by the model at the same IoU threshold.} 
\label{FIG:Confusion}
\end{figure}

In this work, we discuss a deep learning method that uses class labels to obtain instance segmentation masks. 
To evaluate the performance of instance segmentation, we utilize metrics such as Average Precision (AP) and Mean Average Precision (mAP), as shown in Table~\ref{TAB:AP}. 
These metrics assess both the accuracy of detecting specific object classes and the quality of segmentation masks. 
They provide a comprehensive evaluation by considering the ability to localize objects of different classes and the accuracy of their segmentation.
To further evaluate our instance segmentation method, we present confusion matrices that analyze localization and segmentation performance at the object level. 

It is crucial to understand that the confusion matrix for instance segmentation presented here, differs from the one used in classification.
In instance segmentation, an image can contain multiple instances of the same or different classes. 
As a result, the confusion matrix for instance segmentation can have multiple true positive (TP), false positive (FP), and false negative (FN) values for each class.
In contrast, classification assumes only one label per image, resulting in a confusion matrix with only one TP, FP, and FN value for each class.
For instance segmentation, accurately localizing and segmenting object boundaries are vital. 
When computing the confusion matrix, we take into account the spatial overlap between the predicted and ground truth segmentations, which is evaluated using Intersection over Union (IoU). 
This spatial overlap measurement enables us to assess the quality of segmentation and spatial localization performance, which are critical factors in instance segmentation.

Figure \ref{FIG:Confusion} displays the confusion matrices for the training and testing datasets. These matrices focus on instances with an Intersection over Union (IoU) threshold above 0.5.
For easy comparison, the matrix is normalized based on the total number of sources for each class.
The diagonal values represent the TP values. 
In this context, TP refers to instances that are correctly detected and accurately segmented by the model. These instances have an IoU greater than the specified threshold, indicating a strong match with the corresponding ground truth instances.
FP represents instances detected and segmented by the model when, in reality, there is no corresponding ground truth instance. These instances have an IoU greater than the threshold but do not correspond to any true objects.
FN indicates instances that are missed or not detected by the model. These instances are present in the ground truth data but are not identified and segmented by the model, meaning their predicted IoU falls below the specified threshold.

The provided TP, FP, and FN values in the top panel offer insights into the instance segmentation performance of different classes in the training set. 
For the FR-I class, we observe a TP value of 0.74, indicating that the model correctly detected and accurately segmented 74\% of the FR-I source instances in the training set. 
This suggests a reasonably good performance in identifying and aligning with the ground truth FR-I instances. 
However, there is a moderate number of false positives (FP = 0.21), meaning that the model predicted 21\% of instances as FR-I when they did not correspond to any true FR-I instances. 
Additionally, there is a considerable number of false negatives (FN = 0.26), implying that the model missed or failed to detect 26\% of the actual FR-I instances, resulting in a lower IoU score.
Similar patterns can be observed for the FR-II, FR-X, and R classes. 
These classes also demonstrate varying levels of TP, FP, and FN values. 
The FR-II class shows a TP of 0.73, indicating a good rate of correct detection and segmentation, but with a relatively high FP of 0.26 and FN of 0.27. 
The FR-X class exhibits even better performance with a higher TP of 0.81, a lower FP of 0.17, and a lower FN of 0.19. 
Lastly, the R class has a TP of 0.67, suggesting a moderate rate of correct detections but a higher FP of 0.31 and FN of 0.33.

For the test set (bottom panel), the TP, FP, and FN values for FR-I, FR-II and FR-X are relatively consistent with the training set. 
This indicates that the model's performance in detecting and segmenting these source instances remains stable across both datasets.
R class demonstrates a different performance pattern in the test set, as also shown in Table~\ref{TAB:AP} and discussed in Section~\ref{SEC:Results}. 
The TP value is relatively low at 0.45, indicating a moderate rate of correct detections and higher FP and FN values.

In the test set, the TP, FP, and FN values for FR-I, FR-II, and FR-X classes show consistency with the training set, indicating stable performance in detecting and segmenting these sources. 
The R class exhibits a different pattern in the test set, as highlighted in Table~\ref{TAB:AP} and discussed in Section~\ref{SEC:Results}. 
The TP value for the R class is relatively low at 0.45, indicating a moderate rate of correct detections, while the FP and FN values are higher. 
These findings suggest that future studies should focus on potential modifications or refinements to improve the model's performance, particularly for the R class.

\label{lastpage}
\end{document}